\documentclass[pdftex,twocolumn,epjc3]{svjour3}   
\RequirePackage[T1]{fontenc}

\smartqed  

\RequirePackage{graphicx}
\RequirePackage{epstopdf}
\RequirePackage{mathptmx}      
\RequirePackage{flushend}
\RequirePackage[numbers,sort&compress]{natbib}
\RequirePackage[colorlinks,citecolor=blue,urlcolor=blue,linkcolor=blue]{hyperref}

\usepackage{amsmath}
\usepackage{amssymb}
\journalname{Eur. Phys. J. C}

\begin{document}
\title{Integral representation of the Feynman propagators of the Dirac fermions on the de Sitter expanding universe}

\author{Ion I. Cot\u aescu 
\thanks{e-mail: i.cotaescu@e-uvt.ro}}
\institute{West University of Timi\c soara, V. P\^ arvan Ave. 4, RO-300223, Timi\c soara, Romania}
\date{Received: date / Revised version: date}

\maketitle

\begin{abstract}
The propagators of the Dirac fermions  on the expanding portion of the $(1+3)$-dimensional de Sitter spacetime are considered as mode sums in momentum representation with a fixed vacuum of Bunch-Davies type.  The principal  result reported here is a new integral representation of the Feynman propagators of the massive and left-handed massless Dirac fields which can be used for deriving the  Feynman rules of the de Sitter QED in Coulomb gauge or  of an extended  QFT.
\end{abstract}


\maketitle

\section{Introduction}

Important pieces of the quantum field theory on curved spacetimes are the two-point functions that can be calculated either as propagators by using  mode expansions or by looking for new hypotheses complying with the general relativistic covariance as, for example, that of the maximal symmetry of the two-point functions on the hyparbolic spacetimes, i. e. de Sitter (dS) and anti-de Sitter ones \cite{SW}.  

The propagator of the Dirac fermions on the dS spacetime in configuration representation was derived first by Candelas and Reine which  integrated the Green equation for this field \cite{CR}. The same propagator was calculated later as mode sum by Koskma and Prokopec in the context of more general Friedmann-Lema\^ itre-Robertson-Walker spacetimes of arbitrary dimensions \cite{KP}. 

On the other hand, we developed the dS QED in Coulomb gauge \cite{CQED} where we know the Dirac quantum modes in different bases \cite{CD1,CD2,CD3,CdSquant} and we need to use the Feynman propagators for calculating physical effects. However, their expressions as mode sums are not suitable for calculating Feynman diagrams because of their explicite dependence on the Heaviside functions resulted after computing the chronological product. In the flat case this problem is solved by representing the propagators as $4$-dimensional Fourier integrals which include the effects of the Heaviside functions, allowing one to work in momentum representation. Unfortunately, this method cannot be used in the dS case since here the propagators depend on two independent time variables. For this reason,  our objective in this paper  is to find another type of integral representation of the fermion propagators in the $(1+3)$-dimensional dS expanding universe, assuming that the vacuum of the Bunch-Davies type \cite{BuD,BD}  is stable.  

This paper consists of four sections. In the next one we review the fundamental solutions in momentum representation of the Dirac equation minimally coupled to the dS gravity, mentioning that these form a complete system of orthonormalized spinors allowing us to write various mode expansions. For technical reasons, here we express the fundamental solutions in terms of modified Bessel functions \cite{NIST,GR} instead of the Hankel functions used in our previous papers \cite{CD1,CD2,CD3}. In the third section we discuss the propagators of the Dirac field on the dS expanding universe and we propose the principal result of this paper, namely the new  integral representation of the Feynman propagators which encapsulates the effects of the Heaviside functions.  Furthermore, we show that this result is correct since after solving the new integral of this representation we recover the mode sums of the Feynman propagators we know \cite{CD1,KP}.   Some conclusions are presented in the last section.

\section{Fundamental spinor solutions}

Let us first revisit some basic properties of the fundamental solutions of the  Dirac equation minimaly coupled to the gravity of the $(1+3)$-dimensional de Sitter expanding universe. In what follows we consider the normalized solutions of positive and negative frequencies of the momentum-spin basis \cite{CD3} since those of the momentum-helicity basis \cite{CD1} are not defined in rest frames. 

We denote by $M$  the de Sitter expanding universe of radius $\frac{1}{\omega}$ where the notation $\omega$ stands for its Hubble constant. We choose the chart $\{x\}=\{t,\vec{x}\}$ of  {conformal} time, $t\in (-\infty,0]$, and Cartesian coordinates, we refer here as the {\em conformal} chart. This covers the expanding portion of the de Sitter manifold having the line element
\begin{equation}
ds^{2} =\frac{1}{(\omega t)^2}\,\left(dt^{2}- d\vec{x}\cdot
d\vec{x}\right)\,.
\end{equation}
In this chart we consider the non-holonomic local frames defined by the tetrad
fields which have only diagonal components,
\begin{equation}\label{tt}
e^{0}_{0}=-\omega t\,, \quad e^{i}_{j}=-\delta^{i}_{j}\,\omega t \,,\quad \hat
e^{0}_{0}=-\frac{1}{\omega t}\,, \quad \hat e^{i}_{j}=-\delta^{i}_{j}\,
\frac{1}{\omega t}\,.
\end{equation}

In this tetrad-gauge, the massive Dirac field $\psi$ of mass $m$  satisfies the field equations $(D_x-m) \psi (x)=0$  given by the Dirac operator 
\begin{equation}\label{ED}
D_x=-i\omega t\left(\gamma^0\partial_{t}+\gamma^i\partial_i\right)
+\frac{3i\omega}{2}\gamma^{0}\,.
\end{equation}
The general solutions of this equation  may be expanded in terms of fundamental spinors of positive and negative frequencies derived in various representations. Here we consider the momentum representation where the plane wave solutions  $U_{\vec{p},\sigma}$  and  $V_{\vec{p},\sigma}$ depend on the momentum $\vec{p}$ and an arbitrary polarization $\sigma$. These spinors  form an orthonormal basis satisfying the orthogonality relations
\begin{eqnarray}
&&\langle U_{\vec{p},\sigma}, U_{{\vec{p}\,}',\sigma'}\rangle=
\langle V_{\vec{p},\sigma}, V_{{\vec{p}\,}',\sigma'}\rangle =
\delta_{\sigma\sigma^{\prime}}\delta^{3}(\vec{p}-\vec{p}\,^{\prime})\label{U}\\
&&\langle U_{\vec{p},\sigma}, V_{{\vec{p}\,}',\sigma'}\rangle=
\langle V_{\vec{p},\sigma}, U_{{\vec{p}\,}',\sigma'}\rangle =0\,, \label{V}
\end{eqnarray}
with respect to the relativistic scalar product \cite{CD1}
\begin{eqnarray}
\langle \psi, \psi'\rangle&=&\int d^{3}x
\sqrt{|g|}\,e^0_0\,\bar{\psi}(x)\gamma^{0}\psi(x)\nonumber\\
&=&\int d^{3}x
(-\omega t)^{-3}\bar{\psi}(x)\gamma^{0}\psi(x)\,, 
\end{eqnarray}
where $g={\rm det}(g_{\mu\nu})$ while the notation $\bar{\psi}=\psi^+\gamma^0$ stands for the Dirac adjoint of the field  $\psi$. Moreover, this basis is complete satisfying   \cite{CD1}
\begin{eqnarray}
&&\int d^{3}p
\sum_{\sigma}\left[U_{\vec{p},\,\sigma}(t,\vec{x}\,)U^{+}_{\vec{p},\sigma}(t,\vec{x}\,^{\prime}\,)\right. \nonumber\\
&&\hspace*{13mm}\left.+V_{\vec{p},\sigma}(t,\vec{x}\,)V^{+}_{\vec{p},\sigma}(t,\vec{x}\,^{\prime}\,)\right]=(-\omega t)^3\delta^{3}(\vec{x}-\vec{x}\,^{\prime})\,.\label{complet}
\end{eqnarray}

In  momentum representation under consideration here the Dirac field may be expanded as 
\begin{eqnarray}
\psi(t,\vec{x}\,) & = &
\psi^{(+)}(t,\vec{x}\,)+\psi^{(-)}(t,\vec{x}\,)\nonumber\\
& = & \int d^{3}p
\sum_{\sigma}[U_{\vec{p},\sigma}(x)a(\vec{p},\sigma)
+V_{\vec{p},\sigma}(x)b^{\dagger}(\vec{p},\sigma)]~,\label{p3}
\end{eqnarray}
assuming that the particle $(a,a^{\dagger})$ and antiparticle ($b,b^{\dagger})$
operators  satisfy the canonical anti-commutation relations \cite{CD1,CdSquant}, 
\begin{eqnarray}
\{a(\vec{p},\sigma),a^{\dagger}(\vec{p}\,\,^{\prime},\sigma^{\prime})\}&=&
\{b(\vec{p},\sigma),b^{\dagger}(\vec{p}\,\,^{\prime},\sigma^{\prime})\}\nonumber\\
&=&\delta_{\sigma\sigma^{\prime}}
\delta^{3}(\vec{p}-\vec{p}\,^{\prime})\,.
\end{eqnarray}
Thus we obtain a good quantum theory where the one-particle operators conserved via Noether theorem become just the generators of the corresponding isometries \cite{CdSquant}. 

The plane wave solutions can be derived as in Refs. \cite{CD1,CD3} by solving the Dirac equation in the standard representation of the Dirac matrices (with diagonal $\gamma^0$). Here we express these solutions in terms of modified Bessel functions  $K_{\nu}$ \cite{NIST}        
instead of the Hankel functions used in Refs.  \cite{CD1,CD3,CdSquant} since in this manner we get some technical advantages. More specific, working with the real functions $K_{\nu}(z)$ (of complex variables $\nu$ and $z$) we simplify the calculations involving the complex and Dirac conjugations such that the entire formalism becomes more transparent and intuitive, as we shall see in what follows. 

The fundamental spinor solutions in momentum representation of Ref. \cite{CD3} can be rewritten with new suitable phase factors as 
\begin{eqnarray}
U_{\vec{p},\sigma}(t,\vec{x}\,)&=& \sqrt{\frac{p}{\pi\omega}}\,(\omega t)^2\left(
\begin{array}{c}
K_{\nu_{-}}(ip t) \,
\xi_{\sigma}\\
K_{\nu_{+}}(ip t) \,
 \frac{\vec{p}\cdot\mathbf{\sigma}}{p}\,\xi_{\sigma}
\end{array}\right)
\frac{e^{i\vec{p}\cdot\vec{x}}}{(2\pi)^{\frac{3}{2}}}\label{Ups}\\
V_{\vec{p},\sigma}(t,\vec{x}\,)&=&-\sqrt{\frac{p}{\pi\omega}}\, (\omega t)^2 \left(
\begin{array}{c}
K_{\nu_{-}}(-i p t)\,
\frac{\vec{p}\cdot\mathbf{\sigma}}{p}\,\eta_{\sigma}\\
K_{\nu_{+}}(-i p t) \,\eta_{\sigma}
\end{array}\right)
\frac{e^{-i\vec{p}\cdot\vec{x}}}{(2\pi)^{\frac{3}{2}}}\,,\nonumber\\\label{Vps}
\end{eqnarray}
where $p=|\vec{p}|$ and $\nu_{\pm}=\frac{1}{2}\pm i\mu$, with $\mu=\frac{m}{\omega}$.  The Pauli spinors $\xi_{\sigma}$ and $\eta_{\sigma}= i\sigma_2 (\xi_{\sigma})^{*}$ must be correctly normalized,  $\xi^+_{\sigma}\xi_{\sigma'}=\eta^+_{\sigma}\eta_{\sigma'}=\delta_{\sigma\sigma'}$,  satisfying the completeness condition \cite{BDR}
\begin{equation}\label{Pcom}
\sum_{\sigma}\xi_{\sigma}\xi_{\sigma}^+=\sum_{\sigma}\eta_{\sigma}\eta_{\sigma}^+={\bf 1}_{2\times 2}\,.
\end{equation}
In Ref. \cite{CD1} we considered the Pauli spinors of the momentum-helicity basis whose direction of  the spin projection is just that of the momentum $\vec{p}$. However, we can project the spin on an arbitrary direction, independent on $\vec{p}$, as in the case of the
{\em spin} basis  where the spin is projected on the third axis of the rest frame such that  $\xi_{\frac{1}{2}}=(1,0)^T$ and $\xi_{-\frac{1}{2}}=(0,1)^T$ for particles and $\eta_{\frac{1}{2}}=(0,-1)^T$ and $\eta_{-\frac{1}{2}}=(1,0)^T$ for
antiparticles \cite{CD3}. In what follows we work exclusively in this basis called  the momentum-spin basis. 

The form of the spinors (\ref{Ups}) and (\ref{Vps}) suggests us to introduce the auxiliary $4\times 4$  matrix-functions
\begin{equation}
W_{\pm}(x)=\left(
\begin{array}{cc}
K_{\nu_{\pm}}(i x)&0\\
0&K_{\nu_{\mp}}(i x)
\end{array}\right)\,,\quad \forall x\in {\Bbb R}\,, 
\end{equation}
which have the obvious properties
\begin{equation}
\bar{W}_{\pm}(x)=W_{\pm}(x)^*=\gamma^5 W_{\pm}(-x)\gamma^5=W_{\mp}(-x)
\end{equation}
and satisfy 
\begin{equation}
Tr\left[W_{\pm}(x)W_{\mp}(-x)\right]=\frac{2\pi}{x}\,,
\end{equation}
as it results from Eq. (\ref{H3}) and observing that ${\rm Tr}(\pi_{\pm})=2$.
With their help we can write the fundamental spinors in a simpler form as
\begin{eqnarray}
U_{\vec{p},\sigma}(t,\vec{x}\,)&=& \sqrt{\frac{p}{\pi\omega}}\,\frac{e^{i\vec{p}\cdot\vec{x}}}{(2\pi)^{\frac{3}{2}}}\,(\omega t)^2\, W_-(pt) \gamma(\vec{p}) u_{\sigma}
\label{Ups1}\\
V_{\vec{p},\sigma}(t,\vec{x}\,)&=&\sqrt{\frac{p}{\pi\omega}}\, 
\frac{e^{-i\vec{p}\cdot\vec{x}}}{(2\pi)^{\frac{3}{2}}}\,(\omega t)^2\,W_-(-pt) \gamma(\vec{p})v_{\sigma}\,,\label{Vps1}
\end{eqnarray}
depending on the nilpotent matrix
\begin{equation}
\gamma(\vec{p})=\frac{\gamma^0 p-{\gamma^i}p^i}{p}\,, 
\end{equation}
and the 4-dimensional rest spinors of the momentum-spin basis
\begin{equation}
u_{\sigma}=\left(
\begin{array}{c}
\xi_{\sigma}\\
0
\end{array}\right)\quad
v_{\sigma}=\left(
\begin{array}{c}
0\\
\eta_{\sigma}
\end{array}\right)
\end{equation}
which allow us to define the usual projector matrices
\begin{equation}
\pi_+=\sum_{\sigma}u_{\sigma}\bar{u}_{\sigma}=\frac{1+\gamma^0}{2}\,,\quad \pi_-=\sum_{\sigma}v_{\sigma}\bar{v}_{\sigma}=\frac{1-\gamma^0}{2}\,,
\end{equation}
that form a complete system ($\pi_+\pi_-=0$ and $\pi_++\pi_-=1$). All these auxiliary quantities will help us to perform easily the further calculations either by using the form 
\begin{equation}
W_{\pm}(x)=\pi_+ K_{\nu_{\pm}}(ix)+\pi_-K_{\nu_{\mp}}(ix)\,,
\end{equation}
and simple rules as, for example, $\gamma(\vec{p})^2=0$, $~\gamma(\vec{p})\gamma(-\vec{p})=2\gamma(\vec{p})\gamma^0$, $\gamma(\vec{p})\pi_{\pm}\gamma(\vec{p})=\pm\gamma(\vec{p})$,
etc., or resorting to algebraic codes on computer.   

The fundamental solutions in the case of $m=0$ (when $\mu=0$) are derived in Ref. \cite{CD1} using  the chiral representation of the Dirac matrices (with diagonal $\gamma^5$)
and the momentum-helicity basis. We note that these fundamental solutions have fixed helicities (i. e. $-1/2$ for particles and $1/2$ for antiparticles) as in the Minkowski spacetime. This is because the massless Dirac equation is conformally covariant and, therefore,  the  massless spinors in the conformal chart $\{x\}$ are just  the Minkowski ones multiplied with the conformal factor $(-\omega t)^{\frac{3}{2}}$.  

\section{Green functions and Feynman propagators}

Let us consider the partial anti-commutators matrix-functions of positive and negative frequencies \cite{CD1},
\begin{equation}
{S}^{(\pm)}(t,t^{\prime},\vec{x}-\vec{x}\,^{\prime}\,)=-i\{\psi^{(\pm)}(t,\vec{x})\,,\bar{\psi}
^{(\pm)}(t^{\prime},\vec{x}\,^{\prime}\,)\}\,,\label{Spm}
\end{equation}
which satisfy the Dirac equation in both sets of variables \cite{CD1, KP}. The total anti-commutator matrix-function \cite{CD1}
\begin{eqnarray}
{S}(t,t^{\prime},\vec{x}-\vec{x}\,^{\prime}\,)&=&-i\{\psi(t,\vec{x}\,),\bar{\psi}(t',\vec{x}\,^{\prime}\,)\}\nonumber\\
&=&{S}^{(+)}(t,t^{\prime},\vec{x}-\vec{x}\,^{\prime}\,)+{S}^{(-)}(t,t^{\prime},\vec{x}-\vec{x}\,^{\prime}\,)\label{Stot}
\end{eqnarray} 
has similar properties and, in addition, satisfy the equal-time condition
\begin{equation}
{S}(t,t,\vec{x}-\vec{x}\,^{\prime}\,)=-i\gamma^0 (-\omega t)^3\delta^{3}(\vec{x}-\vec{x}\,^{\prime})
\end{equation}
resulted from Eq. (\ref{complet}). 

Let us focus now on the Green functions related to the partial or total anti-commutator matrix-functions.  According to the general definitions \cite{BDR}, we introduced in Ref. \cite{CD1} the retarded (R) and advanced (A) Green functions,
\begin{eqnarray}
S_R(t,t',\vec{x}-\vec{x}\,^{\prime}\,)&=&\theta(t-t')S(t,t',\vec{x}-\vec{x}\,^{\prime}\,)\label{SR}\\
S_A(t,t',\vec{x}-\vec{x}\,^{\prime}\,)&=&-\theta(t'-t)S(t,t',\vec{x}-\vec{x}\,^{\prime}\,)\label{SA}
\end{eqnarray}
and  the Feynman propagator,
\begin{eqnarray}
S_{F}(t,t^{\prime},\vec{x}-\vec{x}\,^{\prime})&=&
-i\langle0|T[\psi(x)\bar{\psi}(x^{\prime})]|0\rangle\nonumber\\
 &=&~  \theta(t-t^{\prime})S^{(+)}(t,t^{\prime},\vec{x}+\vec{x}\,^{\prime}\,)\nonumber\\
 &&-\theta(t^{\prime}-t)
S^{(-)}(t,t^{\prime},\vec{x}-\vec{x}\,^{\prime})~.\label{SF}
\end{eqnarray}
These Green functions satisfy the Green equation that in the conformal chart has the form \cite{CD1},
\begin{equation}
(D_x-m)S_{F/R/A}(t,t^{\prime},\vec{x}-\vec{x}\,^{\prime})=(-\omega t)^3 \delta^{4}(x-x\,^{\prime})\,. \label{p8}
\end{equation}
This equation has an infinite set of solutions corresponding to various initial conditions. However, here we restrict ourselves  to study only the $S_R$, $S_A$ and $S_F$ Green functions, called here propagators,  which may be derived as mode sums without solving the Geeen equation.    

The Feynman propagator (\ref{SF}) can be written as a mode sum since, according to Eqs. (\ref{U}) and (\ref{V}),  the anti-commutator matrix-functions can be put in the form,
\begin{eqnarray}
&&i{S}^{(+)}(t,t^{\prime},\vec{x}-\vec{x}\,^{\prime}\,)=\sum_{\sigma}\int d^3p\, U_{\vec{p},\sigma}(t,\vec{x}\,)\bar{U}_{\vec{p},\sigma}(t\,^{\prime},\vec{x}\,^{\prime}\,) \nonumber\\
&&= \frac{\omega^3  (tt')^2}{8\pi^4}\int d^3p\,p\, e^{i\vec{p}\cdot(\vec{x}-\vec{x}')}W_{-}(pt)\gamma(\vec{p})W_{+}(-pt')\,,\label{Splus}\\
&&i{S}^{(-)}(t,t^{\prime},\vec{x}-\vec{x}\,^{\prime}\,)=\sum_{\sigma}\int d^3p\, V_{\vec{p},\sigma}(t,\vec{x}\,)\bar{V}_{\vec{p},\sigma}(t\,^{\prime},\vec{x}\,^{\prime}\,) \nonumber\\
&&= \frac{\omega^3  (tt')^2}{8\pi^4} \int d^3p\,p\, e^{i\vec{p}\cdot(\vec{x}-\vec{x}')}W_-(-pt)\gamma(-\vec{p})W_+(pt')\,,\nonumber\\
 \label{Smin}
\end{eqnarray} 
after changing $\vec{p}\to-\vec{p}$ in the last integral. Hereby we obtain the generic expression of the Feynman propagator that can be studied either in the configuration representation or in the momentum one.

In configuration representation these matrix-functions can be put in a closed form  particularizing the general results of Ref. \cite{KP} to  $D=4$ dimensions and the Bunch-Davies vacuum. In Ref.  \cite{Cpr} we present the details of this calculation for $m\not= 0$ pointing out that in the massless case we obtain a different result for the left-handed fermions (neutrinos). This is because  we do not have a general definition, in any dimensions, of these fields which seem to be specific to the case of  $D=4$.   

The propagators  (\ref{SR}), (\ref{SA}) and (\ref{SF}) cannot be used in the concrete calculations of Feynman diagrams because of their explicite dependence on the Heaviside  $\theta$-functions.  In the case of the Minkowski spacetime this problem is solved by representing these propagators as $4$-dimensional Fourier integrals which  take over the effects of the Heaviside functions according to the well-know method of the contour integrals \cite{BDR}. In this manner one obtains a suitable integral representation of the Feynman propagators allowing one to work in  momentum representation.

In dS spacetimes we also have a momentum representation but we do not know how to exploit it since in this geometry the propagators are functions of two time variables, $t-t'$ and $tt'$, instead of the unique variable $t-t'$ of the Minkowski case. This situations generates new difficulties since apart from a Fourier transform in  $t-t'\in {\Bbb R}$ a suplementary Mellin transform  for the new variable $tt'\in {\Bbb R}^+$ \cite{GR} might be considered. Obviously, an integral with  two more variables of integration is not a convenient solution for representing the Feynman propagators. 

{ \begin{figure}
  \centering
    \includegraphics[scale=0.30]{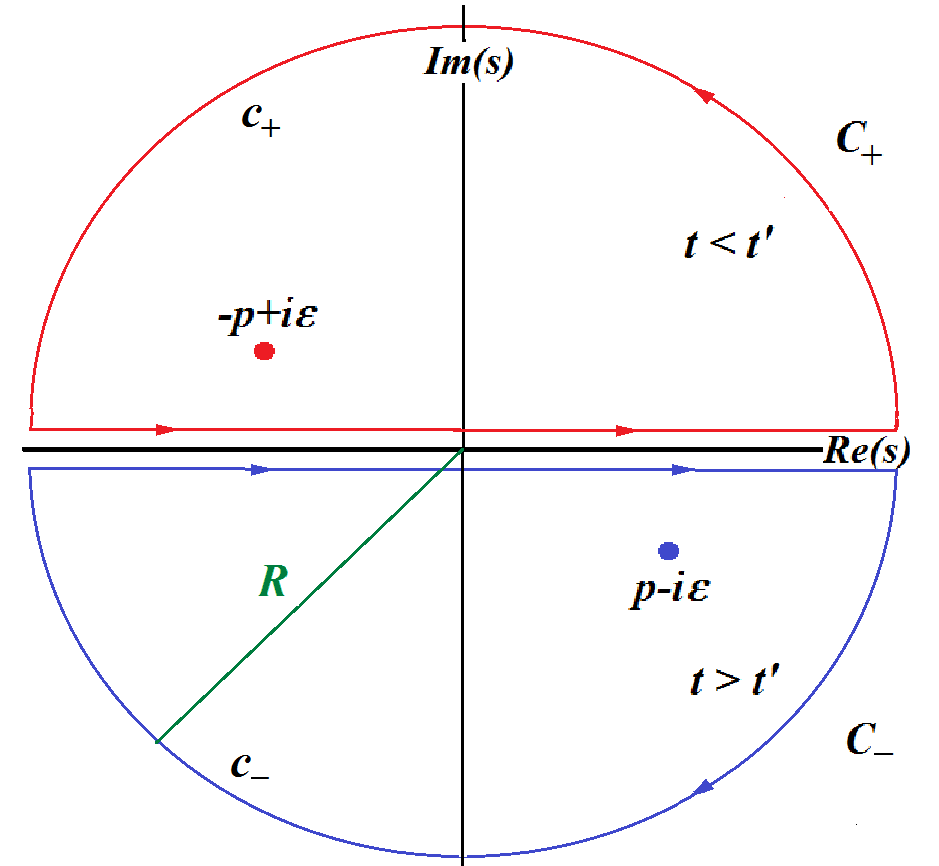}
    \caption{The contours of integration in the complex $s$-plane, $C_{\pm}$, are the limits of the pictured ones for $ R\to \infty$.}
  \end{figure}}

Under such circumstances, we must  look for an alternative integral representation based on  the method of the contour integrals  \cite{BDR} but avoiding the mentioned Fourier or Mellin transforms. The form of the matrix-functions (\ref{Splus}) and (\ref{Smin}) suggests us to introduce a new variable of integration, $s\in{\Bbb C}$, {\em postulating}  the following suitable integral representation of the Feynman propagator of the massive Dirac field,
\begin{eqnarray}
&&{S}_F(t,t^{\prime},\vec{x}-\vec{x}\,^{\prime}\,)
= \frac{\omega^3}{\pi^2} (tt')^2\nonumber\\
&&\times\int {d^3p}\,\frac{e^{i\vec{p}\cdot(\vec{x}-\vec{x}')}}{(2\pi)^3}\int_{-\infty}^{\infty}ds\, |s|\, W_{-}(st)\frac{\gamma^0s-\gamma^i p^i}{s^2-p^2+i\epsilon}W_{+}(-st')\,,\nonumber\\\label{SF1}
\end{eqnarray}     
which encapsulates the effect of the Heaviside functions in a similar manner as in the flat case.  

The main task is to prove that this integral representation gives just the Feynman propagator (\ref{SF})  after solving the integral along the real  $s$-axis that can be written with a self-explanatory notation as 
\begin{equation}
I(t,t')=\int_{-\infty}^{\infty}ds\,M(s,t,t')\,.
\end{equation}
For  large values of $|s|$ we may use the last property of  (\ref{Km0}) obtaining the asymptotic behavior 
\begin{equation}
M(s,t,t')\sim \frac{\gamma^0s-\gamma^i p^i}{s^2-p^2+i\epsilon} \frac{\pi }{2\sqrt{tt'}}\,e^{-is(t-t')}\,, 
\end{equation}
which allows us to estimate the integrals on the semicircular parts, $c_{\pm}$, of the contours pictured in Fig. 1 as,
\begin{equation}
\int_{c_{\pm}}ds\,M(s,t,t')\sim I_0[\pm R(t-t')]\sim \frac{1}{\sqrt{R}}\,e^{\pm R(t-t')}\,,
\end{equation}
according to the first of Eqs. (\ref{Km0}). In the limit of $R\to \infty$ the contribution of $c_+$ vanishes for $t'>t$ while those of $c_-$ vanishes for $t>t'$. Therefore, the integration along the real $s$-axis  is equivalent with the following contour integrals
\begin{equation}
I(t,t')=\left\{
\begin{array}{lll}
\int_{C_+}ds\,M(s,t,t')=I_+(t,t')&{\rm for}& t<t'\\
\int_{C_-}ds\,M(s,t,t')=I_-(t,t')&{\rm for}&t>t'
\end{array}\right. \,,
\end{equation} 
where the contours $C_{\pm}$ are the limits for $R\to \infty$ of those of Fig. 1. Under such circumstances we may apply  the Cauchy's theorem \cite{Complex}, 
\begin{equation}
I_{\pm}(t,t')=\pm 2\pi i \left.{\rm Res}\left[M(s,t,t')\right]\right|_{s=\mp p\pm i\epsilon}\,,
\end{equation}
taking into account that in the simple poles at $s=\pm p\mp i\epsilon$ we have the residues
\begin{eqnarray}
&&\left.{\rm Res}\left[M(s,t,t')\right]\right|_{s=\pm p\pm i\epsilon}\nonumber\\
&&\hspace*{16mm}=\frac{p}{2}W_-(\pm pt)\gamma(\pm\vec{p})W_+(\mp pt')\,.
\end{eqnarray}
Consequently,  the integral $I_-(t,t')$ gives the first term of the Feynman propagator (\ref{SF}) while the integral $I_+(t,t')$ yields its second term. Thus we demonstrated that the integral representation (\ref{SF1}) is correct since after integration over $s$ we obtain just the Feynman propagator (\ref{SF}). 

Note that the other propagators, $S_A$ and $S_R$, can be represented in a similar manner but changing the positions of the poles as in the flat case \cite{BDR}. 

For the left-handed massless fermions the Feynman propagator  can be calculated as  
\begin{equation}
{S}^0_{F}(t,t^{\prime},\vec{x}-{\vec{x}\,}^{\prime})=\lim_{\mu\to 0}\frac{1-\gamma^5}{2}\, {S}_{F}(t,t^{\prime},\vec{x}-{\vec{x}\,}^{\prime})\,\frac{1+\gamma^5}{2}\,,
\end{equation}
taking into account that for $\mu=0$ we may use the particular functions (\ref{Km0}). Thus we arrive to the final result
\begin{eqnarray}
&&{S}^0_{F}(t,t^{\prime},\vec{x}-\vec{x}\,^{\prime}\,)
=\frac{\omega^3}{(2\pi)^4} (tt')^{\frac{3}{2}}\nonumber\\
&&\times\int {d^3p}\int_{-\infty}^{\infty}ds\,\frac{1-\gamma^5}{2}\frac{\gamma^0s-\gamma^i p^i}{s^2-p^2+i\epsilon}\,e^{i\vec{p}\cdot(\vec{x}-\vec{x}')-is(t-t')}\,.\label{SF2} 
\end{eqnarray}
The form of this propagator  written in the conformal chart $\{x\}$ is very similar with that of the flat case because of the conformal covariance of the massless Dirac equation. For this reason the  dS propagator is just the flat one with  the conformal factor $\omega^3 (tt')^{\frac{3}{2}}$.  Obviously, in the flat limit, for $\omega\to 0$ and $\omega t \to-1$, this factor tends to 1 and we recover the neutrino propagator in Minkowski spacetime.

\section{Concluding remarks}

The integral representations of the Feynman propagators we propose here are suitable for calculating Feynman diagrams  where the integration over the supplemental variables $s$ will appear in each internal fermionic line. On the other hand, we know that the contributions of the electromagnetic field is similar as in the Minkowski case since the Maxwell equations are conformally invariant \cite{CQED}. Thus, after solving the space integrals generating $3$-dimensional Dirac $\delta$ functions and integrating over momenta we remain in each diagram with a time integral for each vertex and an integral over the internal variables of the internal lines. Solving all these integrals we may obtain the desired amplitudes in momentum representation. Another advantage of our proposal is that now we can use simple methods of regularization as, for example, the Pauli-Villars one. 

Concluding we may say that now we have all the pieces we need for starting the perturbation machinery of the dS QED in Coulomb gauge which will give us the scattering amplitudes and their radiative corrections in the presence of the dS gravity.

\appendix

\section{Modified Bessel functions}

According to the general properties of the modified Bessel functions, $I_{\nu}(z)$ and $K_{\nu}(z)=K_{-\nu}(z)$ \cite{GR}, we
deduce that those used here, $K_{\nu_{\pm}}(z)$, with
$\nu_{\pm}=\frac{1}{2}\pm i \mu$ are related among themselves through
\begin{equation}\label{H1}
[K_{\nu_{\pm}}(z)]^{*}
=K_{\nu_{\mp}}(z^*)\,,\quad \forall z \in{\Bbb C}\,,
\end{equation}
satisfy the equations
\begin{equation}\label{H2}
\left(\frac{d}{dz}+\frac{\nu_{\pm}}{z}\right)K_{\nu_{\pm}}(z)=-K_{\nu_{\mp}}(z)\,,
\end{equation}
and the identities
\begin{equation}\label{H3}
K_{\nu_{\pm}}(z)K_{\nu_{\mp}}(-z)+ K_{\nu_{\pm}}(-z)K_{\nu_{\mp}}(z)=\frac{i\pi}{ z}\,,
\end{equation}
that guarantees the correct orthonormalization properties of the fundamental spinors. 
Note that for $|z|\to \infty $ and $|{\rm ph}\, z|\leq \frac{\pi}{2}$ we have the asymptotic behaviors 
\begin{equation}\label{Km0}
I_{\nu}(z) \to \sqrt{\frac{\pi}{2z}}e^{z}\,, \quad K_{\nu}(z) \to K_{\frac{1}{2}}(z)=\sqrt{\frac{\pi}{2z}}e^{-z}\,.
\end{equation} 
for any $\nu$ \cite{NIST}.

\end{document}